# Detection of Transformer Winding Axial Displacement by Kirchhoff and Delay and sum Radar Imaging Algorithms

Mohammad S. Golsorkhi, R. Mosayebi, M. A. Hejazi, G. B. Gharehpetian and H. Sheikhzadeh

**Abstract**- In this paper, a novel method for in detail detection of the winding axial displacement in power transformers based on UWB imaging is presented. In this method, the radar imaging process is implemented on the power transformer by using an ultra-wide band (UWB) transceiver. The result is a 2-D image of the transformer winding, which is analyzed to determine the occurrence as well as the magnitude of the winding axial displacement. The method is implemented on a transformer model. The experimental results illustrate the effectiveness of the proposed method.

## I. INTRODUCTION

One of the major causes of transformer failure is the winding mechanical faults [1]. Mechanical faults occur due to insulation aging, high short-circuit currents, transportation lapse, etc. They can gradually weaken the winding insulation capability, which in turn may cause an electrical short circuit inside the transformer and severely damage the transformer winding. Therefore, on time detection of mechanical faults is of high importance.

Concerning transformer winding mechanical fault detection, several methods have been presented in the literature, including the transfer function method [1]-[4], the low voltage impulse method [5], the short circuit impedance method [6], S-parameter measurement method [7] and the more recent ultra wide band (UWB) sensor method [8], [9], [10]. Each of these methods utilizes a characteristic signal, which depends on the geometrical shape of the winding. For each transformer, the characteristic signal is measured and defined as its fingerprint. In case of a mechanical fault the signal will deviate from the transformer's fingerprint. Therefore, the mechanical faults can be detected by comparing the characteristic signal with the fingerprint. In all of these methods, the characteristic signal is a one dimensional (1-D) signal, which provides a limited amount of information about the fault. They are not capable of determining the dimension and type of mechanical faults.

In general, there are two types of mechanical faults, namely radial deformation and axial displacement. They are characterized by a protuberance on the surface of the winding and a change of winding position in vertical direction, respectively. A method for an in-detail detection of the winding radial deformation has been presented in [11]. In this method, a 2-D image of the transformer winding is formed by using radar imaging techniques. By analyzing the resulting image, the position and size of deformation can be detected.

In this paper, a novel method for the detection of the winding axial displacement is presented. It is based on radar imaging methods, which have been utilized for a variety of applications including medical engineering [12], detecting water pipes located inside walls [13] and subsurface sensing [14]. In this method, by using a UWB transceiver, a short pulse is transmitted toward the transformer winding and its reflections are measured. By changing the transceiver antennas the measuring process is repeated along a vertical line on the transformer tank. These measured time signal form a 2-dimensional signal, which is a function of time and antenna position. By using the migration algorithms, this signal is mapped to a 2-D image of the transformer winding [15]. The resulting image graphically represents the geometrical shape of the winding. In case of occurrence of mechanical faults, the resulting image will change. So, the axial displacement appears as a change in the vertical position of the winding image.

In order to verify the effectiveness of the proposed method, an experiment is conducted using a UWB transceiver and a transformer model. The axial displacement is modeled by changing the position of the winding model along its axis. The radar imaging process is implemented with and without winding axial displacement and a 2-D image is formed for each state. In order to determine the exact vertical position of the winding model (the displacement), the resulting images are analyzed by using image processing methods. The magnitude of the axial displacement is calculated by comparing the vertical position of the winding in the resulting images. The experimental results show that the magnitude of the calculated axial displacement in the resulting image matches with the real axial displacement of the model.

M. S. Golsorkhi is with the Department of Electrical and Computer Engineering, Isfahan University of Technology, Isfahan, Iran. (golsorkhi@iut.ac.ir). R. Mosayebi, G. B. Gharehpetian and H. Sheikhzadeh are with Department of Electrical Engineering, Amirkabir University of Technology, Tehran, Iran. M. A. Hejazi is with Department of Electrical Engineering, University of Kashan, Kashan, Iran.

## II. UWB IMAGING

### A. Data recording

The basic principle of UWB radars is demonstrated in Fig. 1. The transmitter generates a short pulse, which propagates through the environment. When the transmitted signal reaches the target, a fraction of its energy reflects back to the receiver. The received signal includes a short pulse, which has a time delay proportional to the distance between the transmitter, target and receiver. Therefore, the presence of the target at a specific distance can be detected.

In order to gain more information about the target, the antennas are moved along a linear path to measure and record the target reflection at different points. Each recorded signal, which is a function of time, is considered as a scan of the target. The set of scans constitute a 2-D signal, which is a function of time and the antennas position. This signal is processed to take a 2-D image of the target [16].

Fig. 2 depicts the concept of the UWB imaging on the transformer winding. Here, the transformer winding is the target. The transmitting antenna (TX) and receiving antenna (RX) are mounted on a horizontal leg. The antennas are moved along the X-axis, and the recording process is conducted at several points. The distance between adjacent measuring points is a constant, which is determined based on the bandwidth of the transmitted pulse. In order to increase the image resolution, the recording span, i.e., the distance between first and last measuring points, should be as large as possible.

Note that the winding of a transformer is located inside a tank, which is filled by transformer oil. Therefore, the antennas should be placed out of the tank. On the other hand, since the tank is constructed from conductor materials, the electromagnetic wave cannot penetrate through it. In order to implement UWB imaging on a transformer, a rectangular section of the tank can be cut and filled by insulating materials. The antennas are placed behind this insulator "window" and the UWB imaging process can be conducted. This technique has been used for the partial discharge monitoring of transformers by UHF sensors [17].

The resolution of the resulting image in the X-axis direction (Δx) is calculated, as follows [11]:

$$\Delta x = \frac{c}{f_0} \cdot \frac{P}{2L} \qquad (1)$$

where, $\lambda_0 = c/f_0$ is the carrier wavelength, c, $f_0$ is the wave speed and UWB pulse center frequency, respectively. L is the measuring range on the X-axis and P is the distance between X-axis and the target.

### B. Data Preprocessing

When the transmitted pulse propagates through the transformer tank, it meets different objects, and a fraction of its energy reflects back to the transmitting antenna. Therefore, a number of short pulses with different delays are recorded by the receiver. In order to take an appropriate image of the winding, the winding reflection should be extracted from the received signal by using a proper time window [11].

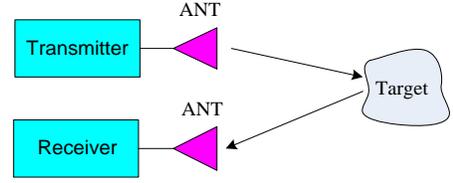

Fig. 1. Basic principle of UWB radar

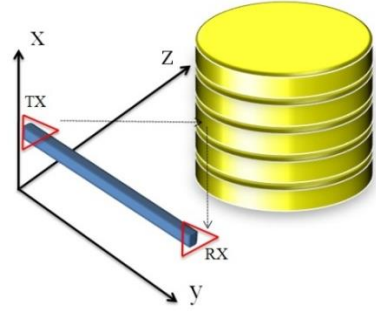

Fig. 2. UWB imaging on transformer winding

It is worthwhile to mention that in case of a real power transformer, the transmitted wave hits the transformer tank several times and multiple reflections appear in the received signal. Fortunately, since the dimensions of high voltage transformers are in the order of meter and the wavelength of UWB pulse is in the order of centimeters, the reflections from the winding can be easily discriminated from the tank reflections and extracted by choosing a proper time window. Therefore, extracting the winding reflection is of prominent importance.

## III. MIGRATION ALGORITHMS

Each preprocessed scan is a discrete-time function, which can be shown by a vector. The set of scans can be represented by a matrix, each row of which is a scan. The scans matrix is a function of time, $t$, and antennas abscissa, $x_{ant}$. The process of obtaining an image from the matrix of scans is called migration. Several migration algorithms have been presented in the literature [18]. In this paper, two migration algorithms are presented: The Kirchhoff migration algorithm, which has been used for the detection of winding radial deformation in [11] and the delay and sum beam-forming, which has been utilized for early detection of breast cancer [20],[21].

### A. Kirchhoff migration

When the incident wave meets the target surface, a portion of it reflects back. Therefore, each point of the target surface can be considered as a wave source, emitting a UWB pulse at the zero time, i.e., when the incident wave reaches that point.

Kirchhoff migration algorithm presents a method to back propagate the measured wave field on the boundary of a medium to the field source. According to the Kirchhoff migration theorem, if the wave field along the surface z=0      (

$\Psi(r',t)$) is known the wave emitted from each point of the source can be calculated, as follows [19]:

$$\Psi(r) = \frac{1}{2\pi} \frac{\partial}{\partial z} \int_{z'=0} \frac{\Psi(r', 2R/c)}{R} da' \quad (2)$$

where, r and r' represent the coordinates of the source and the measuring point on the plane z=0 and R is the distance from the source to the measuring point. It is worthy to mention that since the coordinate of the source points (i.e., the target) is not known, the above integral is calculated over each point of the space below z=0. The result will be a finite number on the source points (target surface) and zero elsewhere. Therefore, the target shape can be reconstructed by plotting the function $\Psi(r)$.

Although the derivation of (2) is based on the general 3-D space, it can be modified for the 2-D case by eliminating the dimension "Y". In this case, the plane $z=0$ is diminished to the X-axis and the surface integral is replaced by a line integral, we have:

$$\Psi(x,z) = \frac{1}{2\pi} \frac{\partial}{\partial z} \int \frac{\Psi(x', 2R/c)}{R} dx' \quad (3)$$

Another form of (3) can be derived by applying the normal derivative in the right side of (3). By introducing the dummy variable $t' = 2R/c$ and some manipulation, we have:

$$\Psi(x,z) = \frac{1}{2\pi} \int \frac{z}{R^2 \cdot c} \left[ \frac{\partial}{\partial t'} \Psi(x',t') - \frac{c}{R} \Psi(x',t') \right] \bigg|_{t'=2R/c} dx' \quad (4)$$

In order to calculate this integral, the electric field all over the X-axis should be known. However, the measured scans provide us the electric field over a limited number of points over the X-axis. Therefore, the integral should be approximated by a summation, as follows:

$$\Psi(x,z) = \frac{1}{2\pi} \times \sum_{k=1}^{K} \frac{z \cdot d_M}{R^2 \cdot c} \left[ \frac{\partial}{\partial t'} \Psi(x_k,t') - \frac{c}{R} \Psi(x_k,t') \right] \bigg|_{t'=2R/c} \quad (5)$$

where, $x_k$ and $d_M$ and are the abscissa of $k^{th}$ measuring point and the distance between adjacent measuring points, respectively.

As shown in Fig. 3, in the Kirchhoff migration algorithm, a rectangular area enclosing the target (winding) is chosen and is splitted to several square shaped pixels. Then, for each pixel (P), $\Psi(x,z)$ is calculated by (5). Finally, the image of the target is obtained by plotting $\Psi(x,z)$ over the $X$-$Z$ plane.

*B. Delay and sum beam-forming*

Delay and sum (DAS) beam-forming is a well-known method and has been widely used for breast cancer detection by UWB imaging [22]. This method forms a 2-D image based on the energy distribution of the imaging plane. After pre-processing steps, i.e., time-zero alignments of scans and the windowing approach for separating the target signal from the other parts of the received signal, the amount of energy, which has been reflected from each focal point in the image must be calculated. Since, the received signals have different time-delays (because of different paths between each antenna and each focal point in the image), each of received and pre-processed signals must be time-aligned with respect to each focal point. This will cause the signals to be coherently processed, i.e. it removes the phase difference among received signals.

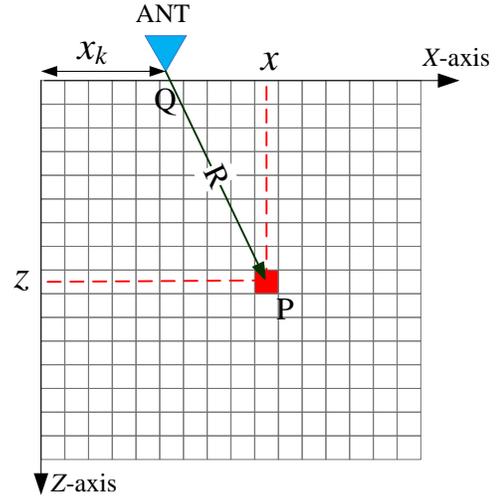

Fig.3. Implementation of Kirchhoff migration

The time-shift required for aligning the signals with respect to each focal point (at location $r_o$) is given by the following equation:

$$n_i(r_o) = \frac{1}{\Delta t} \left\lfloor \frac{\|r_{iT} - r_o\|}{c} + \frac{\|r_{iR} - r_o\|}{c} \right\rfloor \quad (6)$$

where $\lfloor x \rfloor$ stands for rounding to the greatest integer less than $x$, $\|.\|$ denotes the Euclidean norm, $c$ is the approximate velocity of the microwave propagation in the vacuum, $\Delta t$ is the sampling interval and $r_{iT}$, $r_{iR}$ are the transmitter and receiver locations, respectively.

After time-alignment, the signals are coherently processed and the energy distribution of each focal point is calculated by integrating over the signal obtained by the summation of time-aligned signals. This summation results in more information of the target and then the integration gives the energy of the resulting signal. This integration is over a short time-duration of the resulting signal which has been experimentally adjusted. We have [23]:

$$I(\vec{r_0}) = \int_0^L \sum_i^M y_i(n_i(\vec{r_0})) \quad (7)$$

where, $y_i(r_0)$ is the preprocessed signal at each receiver and $n_i(\vec{r_0})$ is the time delay computed for the alignment. The index $i$ represents the transmitter and receiver, and $M$ is the number of measurement points.
Therefore, the DAS beam-forming algorithm can be summarized, as follows:
**Step 1:** The received signals are pre-processed, i.e., the time-zero is calculated and the signals are shifted with respect to it. Then, the target signals must be separated from other parts of received signals.

**Step 2:** The time-gated signals are time-aligned respect to each focal point to process coherently.

**Step 3:** The signals in step 2 are summed.

**Step 4**: An integration of length $L$ is performed over the signal in previous step.

**Step 5:** Steps 2 to 4 are repeated for each focal point to form the matrix $I(\vec{r_0})$.

**Step 6:** The negative values of elements of matrix $I(\vec{r_0})$ are ignored and the resulting matrix is shown as a RGB image [24].

## IV. EXPERIMENTAL RESULTS

Since, a power transformer has a complex structure and considerable dimensions, it is difficult to model it in detail. Therefore, a rather simple model of one phase of the transformer winding, show in Fig. 4, is used in this paper [25]. It consists of a polycarbon cylinder covered by copper strips on its surface. These strips resemble the outer layer of the transformer winding. Since the electromagnetic wave cannot penetrate into metallic transformer winding, it is not necessary to model the inner parts and the iron core. The axial displacement of the winding is modeled by moving the winding in the axial direction [26].

The experimental set-up is depicted in Fig. 5. As shown in this figure, instead of moving the antenna in the vertical direction, the winding is mounted horizontally and the antennas are moved along a horizontal line parallel to the winding axis. A UWB transceiver is utilized to perform the UWB imaging procedure. It consists of a transmitter, a receiver and two Vivaldi type micro-strip antennas. The transmitter and receiver are connected to a PC via Ethernet cable. They can be controlled from the PC by using software provided by the manufacturer. The specifications of the experimental set-up are listed in table 1.

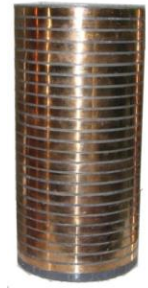

Fig. 4. Winding model

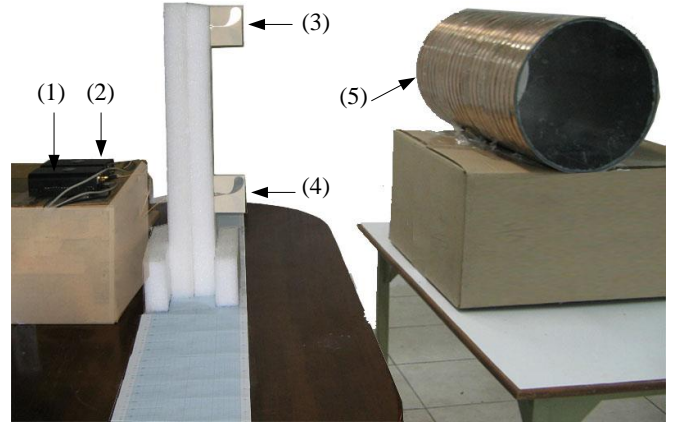

Fig. 5. UWB imaging set-up: 1-transmitter 2-receiver 3-RX antenna 4-TX antenna and 5-transformer model.

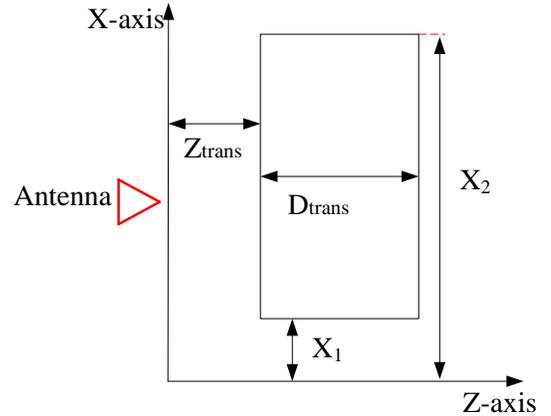

Fig. 6. Top-view of set-up

Table 1. Specifications of experimental set-up

| Parameter | Value | Symbol |
|---|---|---|
| Transceiver pulse repetition frequency | 9.6MHz | $RPF$ |
| UWB pulse center Frequency | 4.7GHz | - |
| UWB pulse bandwidth(-10dB) | 3.2GHz | - |
| Time interval between scans | 20ms | $T_{pulse}$ |
| Distance between antennas | 400mm | $D_{sep}$ |
| Transmit antenna height | 250mm | $H_{tx}$ |
| Receive antenna height | 650mm | $H_{rx}$ |
| Model cylinder height | 650mm | - |
| Model diameter | 300mm | $D_{trans}$ |
| Model height from table | 300mm | |
| Model distance from X-axis | 600mm | $Z_{trans}$ |
| Measuring step | 20mm | $d_M$ |
| Number of measuring points | 60 | $K$ |

The measuring points are marked on a scaled paper on the table. For each measurement, the antennas leg is carefully placed at the measuring point and 100 scans are recorded. These scans are averaged to form the final scan for each measuring point.

Fig. 6 shows the top view of the set-up. In this figure, the model is shown by a rectangle and the abscissa of its lower and upper sides are shown by $X_1$ and $X_2$, respectively. The resultant radar image of the model is expected to match this figure.

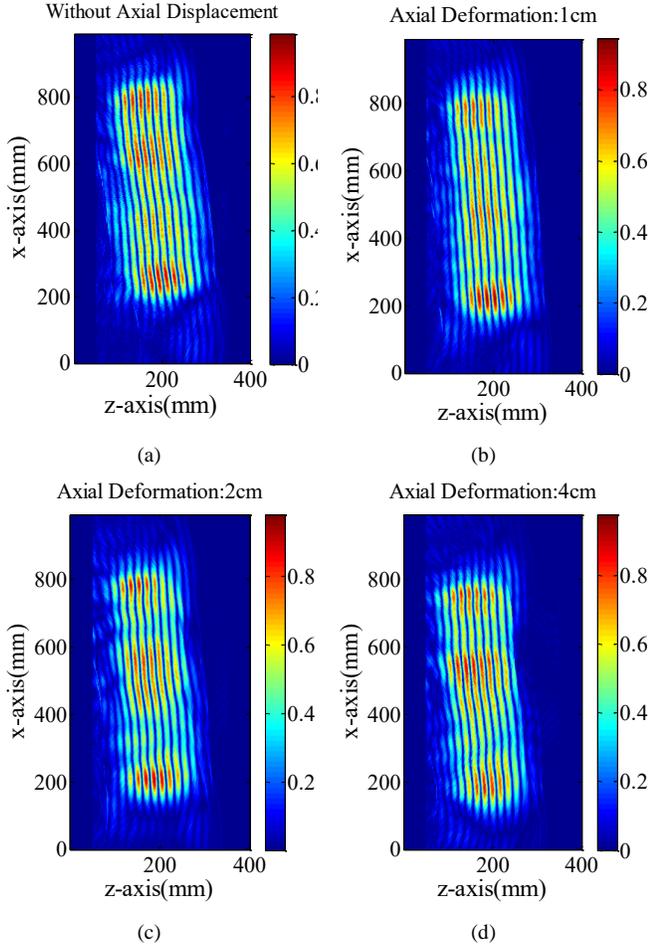

Fig. 7. Radar images of model obtained by Kirchhoff migration algorithm for: a) 0cm, b) 1cm, c) 2cm and d) 4cm displacement.

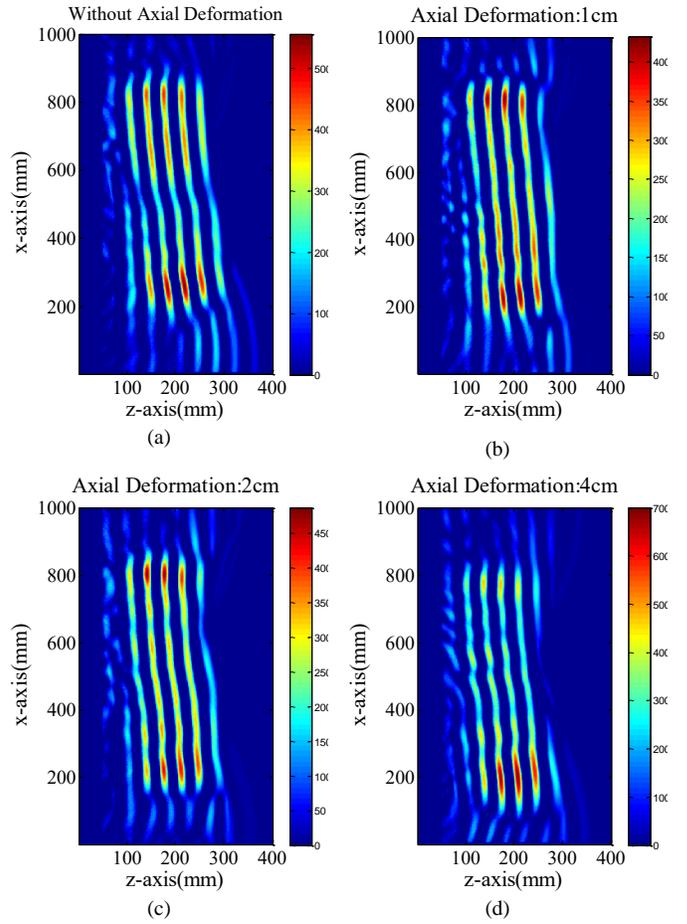

Fig. 8. Radar images of model obtained by DAS algorithm for: a) 0cm, b) 1cm, c) 2cm and d) 4cm displacement.

In order to assess the effectiveness of the proposed method, four different states are studied. In the first state, the model is assumed to be intact, i.e., without axial displacement. In the second, third and fourth states, the model is displaced 10, 20 and 40mm in the X-axis direction, respectively. This displacement resembles the winding axial displacement in power transformers. For each state, the UWB imaging procedure, including data recording, preprocessing and migration is carried out. Then, $\Psi(x,z)$ is plotted in color-map scale, in which the color of each pixel indicates the magnitude of $\Psi$ at that pixel. In this type of plot, cold colors represent a low value, while warm colors show a high value.

The resulted images of the model obtained by Kirchhoff migration beam forming algorithm is shown in Fig. 7. Fig. 7.*a* depicts the radar image of the model for the first state (without axial displacement). The radar images of the model in the second, third and fourth states are shown in Fig. 7.*b*, *c* and *d*, respectively. In each image, the model appears as a number of vertical strips. Although these strips might not seem to match the shape of the model, they bear some important information about it, namely the upper and lower edges of them indicate the upper and lower sides of the model. In other words, the vertical position of the model in each state can be extracted from these images. The axial displacement of the model is calculated by subtracting its vertical position in each state from that of the intact state.

The resulted images of the model obtained by DAS algorithm for the four states are shown in Fig. 8. They are similar to the resulted images of the Kirchhoff migration method, except that in case of Kirchhoff method the number of vertical strips is twice. The vertical position of the model can be extracted from the radar images either manually, i.e., by zooming the images and finding the position of lower and upper edges, or by utilizing image processing techniques. Since the vertical strips in the images fade gradually, the former can be both imprecise and time consuming. Therefore, in this paper, we utilize image processing techniques to efficiently determine the position of the model.

## V. RESULTS PROCESSING

First of all, in order to distinguish the strips, the image is filtered by "open" operator, which is one of the morphology operators. It is defined, as follows [27]:

$$\Psi_B = \left[\Psi(-B)\right](+B) \qquad (8)$$

where, + and – are the symbols of dilation and erosion, which are performed on the image ($\Psi$) by using the structural element ($B$). Fig. 9.*a* and *b* illustrate the original and filtered images of the model, respectively.

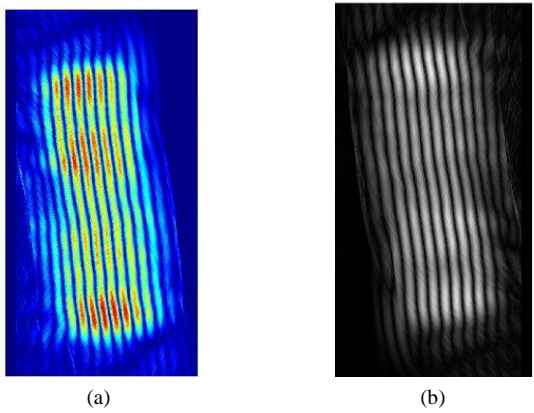

(a) (b)

Fig. 9. Application of open operator on model image

In the second step, the strips are detected by using global thresholding method [28]. During this process, each pixel of the image is defined as "object" pixels if its value is greater than the threshold value and as "background" pixel otherwise. Therefore, the result is a binary image. The threshold value can be selected based on various algorithms. In this paper, the Ostu segmentation method [29] is used, to calculate the optimum threshold separating the strips in the image from the background. The resulting image is depicted in Fig. 10.*a*. In this image, the strips can be clearly distinguished. The short strips are undesirable and are removed as shown in Fig. 10.*b*.

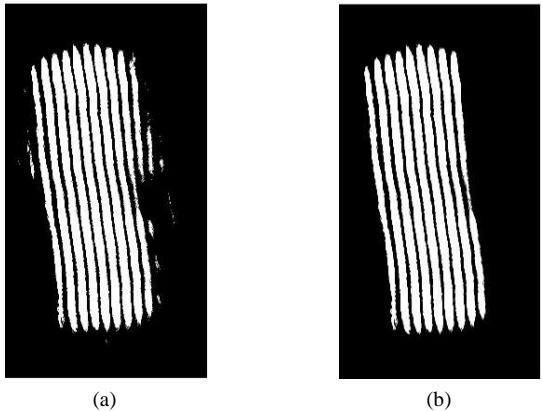

(a) (b)

Fig. 10. Resulting image of model after: a)application of global thresholding method and b)removing short strips

In the third step, one of the strips is chosen as a reference and the other parts of the image are removed, as shown in Fig. 11.*a*. The outer edge of the reference strip is defined by the points at which the image function is equal to a threshold value, as depicted in Fig. 11.*b*. The maximum and minimum abscissas of the edge are taken as the upper and the lower edge of the model, respectively. In Fig. 11.*c* the upper and lower edges of the strip are shown by two red lines.

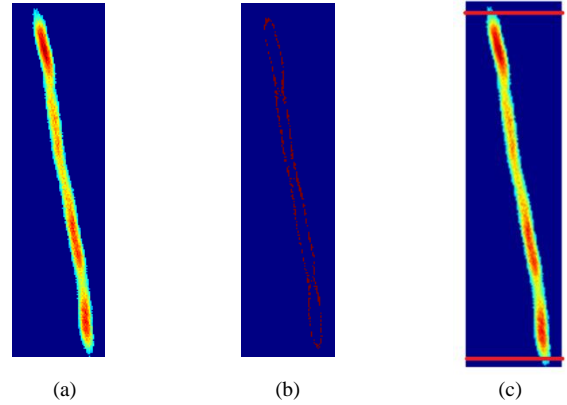

(a) (b) (c)

Fig. 11. Detecting the upper and lower edges

The image processing algorithm can be summarized, as follows:
1. Filtering the original image by open operator (Fig. 9.b)
2. Extracting the vertical strips by global thresholding method (Fig. 10.a)
3. Removing the short strips (Fig. 10.b)
4. Selecting one strip as reference (Fig. 11.a)
5. Extracting the strip edge (Fig. 11.b)
6. Determining the maximum and minimum abscissa of the strip (Fig. 11.c)

The abscissa of the center point of the model is calculated, as follows:

$$X_C = \frac{X_1 + X_2}{2} \quad (9)$$

$X_C$ is a reference for measuring the abscissa of the model. Then the axial displacement of the model in each state is calculated by subtracting its $X_C$ from $X_C$ of the intact case. Finally, the error of the proposed method in each state is calculated by comparing the measured value of the axial displacement with its actual value.

The abscissas of the lower and upper edges of the resulting images of the model obtained by Kirchhoff migration and DAS algorithm are summarized in tables 2 and 3, respectively. In these tables, E is associated to the values estimated by UWB imaging algorithms and A indicates the actual values, measured by a ruler. It is obvious that the estimation error obtained via DAS algorithm is remarkably less than the estimation error obtained via Kirchhoff algorithm. Nevertheless, DAS suffers from computational burden and takes more time to form the image.

Table 2. Experimental results: Kirchhoff algorithm

| Parameter | Des. | Value in state: 1 | 2 | 3 | 4 | Unit |
|---|---|---|---|---|---|---|
| Lower edge abscissa | E | 208 | 192 | 181 | 157 | mm |
| Upper edge abscissa | E | 844 | 837 | 821 | 801 | mm |
| Center abscissa | E | 526 | 514.5 | 501 | 479 | mm |
| Axial displacement | E | - | 11.5 | 25 | 47 | mm |
| Axial displacement | A | - | 10 | 20 | 40 | mm |
| Estimation error | - | - | 15 | 25 | 17.5 | % |

Table 3. Experimental results: DAS algorithm

| Parameter | Des. | Value in state: 1 | 2 | 3 | 4 | Unit |
|---|---|---|---|---|---|---|
| Lower edge abscissa | E | 104 | 84 | 79 | 35 | mm |
| Upper edge abscissa | E | 759 | 755 | 743 | 703 | mm |
| Center abscissa | E | 431.5 | 419.5 | 411 | 369 | mm |
| Axial displacement | E | - | 11 | 20.5 | 42.5 | mm |
| Axial displacement | A | - | 10 | 20 | 40 | mm |
| Estimation error | - | - | 10 | 2.5 | 6.25 | % |

## VI. DISCUSSION

In this paper, a simplified model of three phase transformer has been utilized to verify the effectiveness of the proposed method. In case of a real power transformer, there are two issues to be considered:

1- The axial displacement might be a total displacement of the winding in vertical direction or it can be in the form of disk space variation [28], [29]. In this research the focus is on total displacement of the winding. But detection of disk space variation in the outer layer of the winding is theoretically possible with this method and can be tested and discussed in another paper.

2- In this paper only one side of a one phase winding has been inspected for mechanical faults. However, the full circumference of each winding can be captured by installing one vertically movable antenna set on the front side and another one on the back side of each of the three phase windings, as shown in Fig. 12.

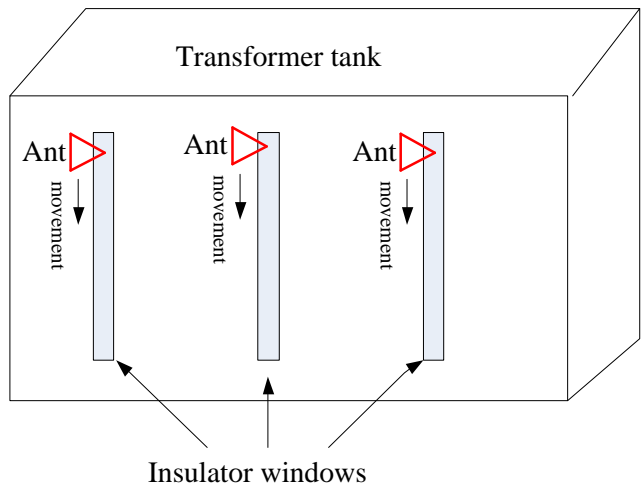

Fig. 12. Implementation of the proposed method on a three phase transformer

## VII. CONCLUSION

A novel method for detecting and measuring the magnitude of the transformer winding axial displacement based on radar imaging has been presented. Two radar imaging algorithms, namely Kirchhoff migration and DAS beam-forming have been implemented using an experimental set-up, which consists of a UWB transceiver and a transformer winding model. The axial displacement has been modeled by moving the model in axial direction. Four experiments have been conducted with zero, 1cm, 2cm, and 4cm axial displacement, and the radar images of each experiment has been extracted by using Kirchhoff and DAS algorithms. The resulting images have been further processed to determine the magnitude of the axial displacement for each state. The experimental results demonstrate that although both methods are effective in detecting the axial displacement, DAS method can estimate the magnitude of the displacement more precisely.